\documentclass[prl,showpacs,floatfix,footinbib,twocolumn]{revtex4}
\usepackage{graphicx}
\usepackage{dcolumn}
\usepackage{epsfig}
\usepackage{float}
\usepackage{bm}

\usepackage{bibentry}

\begin{document}

\bibliographystyle{apsrev}

\title{Multiple-source optical spectral weight transfer in ferromagnetic EuB$_{6}$}

\author{Jungho Kim$^{1}$,
%C. C. Homes$^{2}$,
B. K. Cho$^2$, Young-June Kim$^{1}$, and E. J.
Choi$^{3,}$}\altaffiliation{Corresponding author: echoi@uos.ac.kr}

\affiliation{$^1$Department of Physics, University of Toronto,
Toronto, Ontario M5S 1A7, Canada}
%\affiliation{$^2$ Department of
%Physics, Brookhaven National Laboratory, Upton, NY 11973-5000}
\affiliation{$^2$Center for Frontier Materials and Department of
Materials Science and Engineering, K-JIST, Gwangju 500-712, Korea}
\affiliation{$^3$Department of Physics, University of Seoul, Seoul
130-743, Republic of Korea}

%
% The abstract
%

\begin{abstract}
We present the first wide range (2~meV$-$5.5~eV) optical
conductivity of EuB$_6$ from reflectivity and ellipsometry
measurements. Upon the ferromagnetic transition at T$_{c}$=15.5~K,
interband transition $\sigma_{1}(\omega)$ decreases at three
different energy $\omega$=0.5, 1.4, 2.75~eV and the lost spectral
weight is transferred to the Drude $\sigma_{1}(\omega)$ at $\omega
<$0.33 eV. We succeeded in explaining this unprecedented
multiple-energy $\sigma_{1}(\omega)$ change using LDA+U calculated
band structure by Kunes and Pickett (Phys. Rev. B, $\textbf{69}$,
165111). Our finding supports strongly that (1) EuB$_6$ is a
semimetal and (2) the exchange-driven band splitting is the primary
source of the drastic $\rho(T)$ and $\omega_{p}^{2}$ changes at T
$<$ T$_{c}$.
\end{abstract}


\maketitle

%
% The main body of the text
%

Interplay of itinerant carrier with localized magnetic moment can
lead to novel magneto-resistance effect in solid. When the moments
align with long range ferromagnetic order, dc-resistivity exhibits a
drastic change as seen in EuO and (Ga,Mn)As. Theories such as the
Kondo-lattice model, Ruderman-Kittel-Kasuya-Yosida (RKKY)
interaction and spin-polaron model propose possible microscopic
mechanism of the carrier-moment coupling in magnetic conductors.

Europium hexaboride EuB$_6$ is a simple cubic compound where dilute
conducting carrier ($n\sim$ 10$^{19}$cm$^{-3}$) coexists with
localized 4f spin of Eu$^{+2}$ (S=7/2). The latter undergoes
ferromagnetic (FM) transition at T$_{c}$=15.5~K where simultaneously
dc-resistivity $\rho$ drops sharply \cite{Guy80} and the infrared
plasma frequency $\omega_p^2$ (=$4\pi e^2n/m$) suddenly increases
\cite{degiorgi}. Such dramatic FM-driven changes were seen in the
manganites like (La$_{0.7}$Ca$_{0.3}$)MnO$_3$ \cite{cmr} whereas
EuB$_6$ being free from complications like lattice distortion and
random substitution offers a cleaner and ideal example of
ferromagnetic metal where the carrier-moment interaction can be
studied in more fundamental level.

In the Kondo-lattice model itinerant electron band is split in FM
phase by the s-f exchange interaction into two spin-polarized
sub-bands. Kreissl and Nolting applied this theory to EuB$_6$ to
find that the band splitting brings about a significant increase of
the carrier density \cite{kreissl}. The $\rho(T)$ drop and
$\omega_p^2$ increase are attributed to it. In contrast Hirsch
proposed an alternative model that conduction band becomes broader
in FM state because the spin-polarization reduces the bond-charge
repulsion effect \cite{hirsch99,hirsch}. This band width increase or
equivalently carrier mass decrease is considered as a source of the
enhanced metallicity. Pereira $et~al.$ suggested yet another
scenario that at T$>$T$_{c}$ the disordered spin background can
localize the carriers via their exchange interaction and it
introduces a mobility edge in the conduction band \cite{pereira}. FM
spin alignment reduces the localization effect, lowering the
mobility edge position, and the carriers become delocalized. This
theoretical diversity shows that EuB$_6$ is an important prototype
of metallic ferromagnetism. At present it is not clear which model
is correctly describing it.

Probing electronic band structure is an essential step to solve this
issue. According to the de Haas-van Alphen and Shubnikov-de Hass
experiments \cite{goodrich,aronson}, electron- and hole-pockets
coexist on the Fermi surface of EuB$_6$. This semimetal picture is
employed in the Kondo-lattice theory of Ref.~4. On the other hand,
angle-resolved photoemission spectroscopy (ARPES) showed that
EuB$_6$ is an insulator with a sizable gap of 1~eV \cite{denlinger}.
The localization theory of Pereira $et~al.$ is constructed based on
this picture. This apparent controversy on the electronic band
structure is by itself another long-standing open question. In
particular none of the experiments mentioned above revealed how the
PM electronic structure changes in FM phase. Clearly current
understanding of this material is far from complete and further
study is needed to resolve the two issues, i.e, origin of FM-driven
changes and nature of the electronic band structure.

In this Letter, we used reflectivity and ellipsometry measurements
to obtain the optical conductivity spectrum of EuB$_6$ over the wide
frequency region 2~meV$-$5.5~eV which greatly extends the previous
infrared work \cite{degiorgi}. We observe that at T$<$T$_{c}$
$\sigma(\omega)$ exhibit rich and novel changes in the interband
transition region up to 4~eV as well as in the Drude region. We
propose an electronic band model that explains all the spectral
changes coherently. It allows simultaneous solutions to the
aforementioned two issues.

%
% Experimental
%
Single-crystal EuB$_6$ samples were synthesized by the boro-thermal
flux method \cite{rhyee2,rhyee3}. The normal incidence reflectivity,
$R(\omega)$, was measured at 5~K$\leq$T$\leq$100~K in the
2~meV$-$1~eV range using a Fourier transform spectrometer with the
\textit{in situ} overcoating technique \cite{homes93}. For
0.7$-$5.5~eV range, optical dielectric constants were measured using
spectroscopic ellipsometor (Woollam VASE).

%results

%
% Figure 1
%
\begin{figure}[t]
\vspace*{-0.0cm}\centerline{%
\includegraphics[width=2.0in,angle=0]{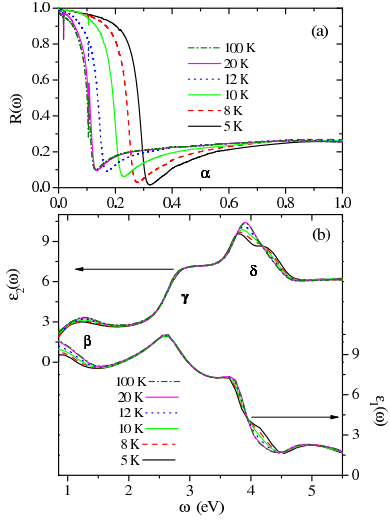}}
%\centerline{\includegraphics[width=4.5in,angle=-90]{fig1.eps}}%
\vspace*{0.0cm}%
%\centering\epsfig{file=fig1.eps,width=5cm,angle=-90}
%
\caption{(Color online) Optical spectra of EuB$_6$ for T=5, 8, 10,
12, 20, and 100~K. (a) Reflectivity at $\omega < $1~eV. (b)
Dielectric constants $\varepsilon_{2}(\omega)$ and
$\varepsilon_{1}(\omega)$ for $0.7 < \omega < $5.5 ~eV from
spectroscopic ellipsometry measurement. The peaks $\beta, \gamma$,
and $\delta$ represent the interband transitions.} \label{fig:fig1}
\end{figure}

Fig.~\ref{fig:fig1}(a) shows $R(\omega)$ of EuB$_6$ for
5~K$\leq$T$\leq$100~K. The high reflectivity in the low frequency
region is due to the metallic Drude response. The plasma edge shifts
to higher energy as T decreases below 20~K, which is consistent with
the earlier report by Degiorgi $et~al.$ \cite{degiorgi}. Complex
dielectric constants $\varepsilon_{2}(\omega)$ and
$\varepsilon_{1}(\omega)$ from the high-frequency ellipsometry
measurement are shown in Fig.~\ref{fig:fig1}(b). Three absorption
peaks are observed at 1.4~eV, 2.8~eV, and 4~eV (labeled as $\beta,
\gamma$, and $\delta$ respectively) which correspond to interband
electronic transitions. The spectra exhibit systematic change with
T. It occurs even at $\omega >$ 4eV which is remarkably higher than
the temperature scale.

\begin{figure}[t]
\vspace*{-0.0cm}\centerline{%
\includegraphics[width=2in,angle=-90]{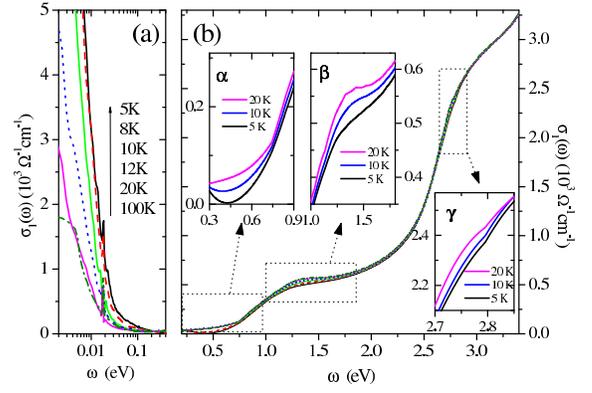}}
%\centerline{\includegraphics[width=4.5in,angle=-90]{fig1.eps}}%
\vspace*{0.0cm}%
%\centering\epsfig{file=fig1.eps,width=5cm,angle=-90}
%
\caption{(Color online) Optical conductivity $\sigma_{1}(\omega)$ of
EuB$_6$ for (a) Drude part at $\omega<$0.33~eV (b) interband (IB)
part at 0.33~eV $<\omega<$3.3~eV. The insets in (b) are close-up
views of the three features $\alpha$, $\beta$ and $\gamma$.}
\label{fig:fig2}
\end{figure}

To investigate the T-dependent change, we extract optical
conductivity $\sigma_{1}(\omega)$ from $R(\omega)$ and
$\varepsilon_{1,2}(\omega)$. We used the Kramers-Kronig (KK)
constrained variational fitting method in this analysis
\cite{kuzmenko05}. Fig.~\ref{fig:fig2} shows $\sigma_{1}(\omega)$ in
the Drude part and in the interband transition (IB) part separately.
At T$<$T$_{c}$ Drude $\sigma_{1}(\omega)$ increases reflecting the
blue-shift of the $R(\omega)$ plasma edge. On the other hand, IB
transition peaks $\beta$ and $\gamma$ show the opposite behavior,
the $\sigma_{1}(\omega)$ decrease. The decrease is seen also at
$\omega\sim$0.5~eV where the IB transition sets in. We label this IB
onset as $\alpha$ \cite{com1}. The peak $\delta$ will be discussed
separately later.

% Figure 2
%

\begin{figure}[t]
\vspace*{-0.0cm}\centerline{%
\includegraphics[width=2in,angle=-90]{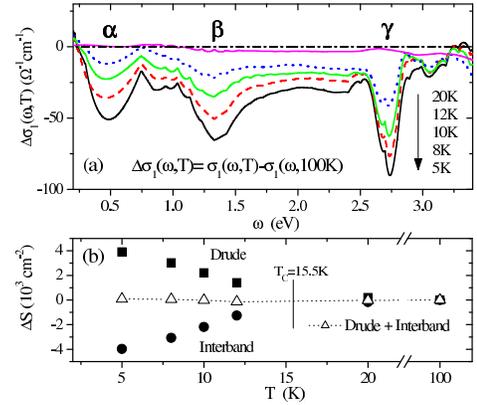}}
%\centerline{\includegraphics[width=4.5in,angle=-90]{fig1.eps}}%
\vspace*{0.0cm}%
%\centering\epsfig{file=fig1.eps,width=5cm,angle=-90}
%
\caption{(Color online) (a) Interband conductivity difference with
temperature
$\Delta\sigma_1(\omega,T)$=$\sigma_1(\omega,T)$-$\sigma_1(\omega,100K)$.
(b) Spectral weight change $\Delta
S$=$\frac{120}{\pi}\int[\Delta\sigma_1(\omega,T)]d\omega$ of the
Drude part (0$<\omega<$0.33~eV), the IB part (0.33$<\omega<$3.3~eV),
and the sum of the two parts (0$<\omega<$3.3~eV).} \label{fig:fig3}
\end{figure}

In Fig.~\ref{fig:fig3}(a), we show the $\sigma_{1}(\omega)$ change
more closely by plotting the difference curve
$\Delta\sigma_1(\omega,T)$=$\sigma_1(\omega,T)$-$\sigma_1(\omega,100K)$.
The three negative humps represent the $\sigma_{1}(\omega)$ decrease
at $\alpha$, $\beta$ and $\gamma$. The humps appear at T=12~K ($<$
T$_{c}$=15.5~K) showing that they are induced by the FM transition.
For quantitative analysis, we calculate the IB spectral weight
change $\Delta S$=$\frac{120}{\pi}\int[\Delta\sigma_1(\omega,T)
]d\omega$ over 0.33$<$$\omega$$<$3.3~eV and also for the
Drude-change at 0$<$$\omega$$<$0.33~eV. Fig.~\ref{fig:fig3}(b) shows
that $\Delta S_{IB}$ decreases while $\Delta S_{Drude}$ increases
below T$_{c}$. Note that they almost cancel each other $\Delta
S_{Drude}$+$\Delta S_{IB}\cong $0. It means that
$\sigma_{1}(\omega)$ weight is transferred from the IB part to the
Drude part. Optical weight transfer is widely seen in strongly
correlated electron materials particularly insulator-metal
transition compounds where Drude weight change comes from some
region at higher frequency. EuB$_6$ is, however, very different in
that the Drude weight comes not from single high-$\omega$ source but
from the three different $\alpha$, $\beta$ and $\gamma$-humps. Such
multiple-source $\sigma_{1}(\omega)$ weight transfer is unique and
unprecedented. Further, while $\alpha$-hump is on border of the
Drude peak, $\beta$- and $\gamma$-hump are far remote from it.
$\gamma$-hump is located at 2.75~eV which evidently has no energy
overlap with the Drude conductivity at $<$0.33~eV. It is remarkable
that nevertheless the two separate features are ``communicating"
with each other. These novel behaviors of interband
$\sigma_{1}(\omega)$ suggest that electronic band structure of
EuB$_6$ undergoes an unusual change at T$_{c}$.

% Figure 3
%

\begin{figure}[tbp]
\vspace*{0.2cm}\centerline{%
\includegraphics[width=2.8in,angle=-90]{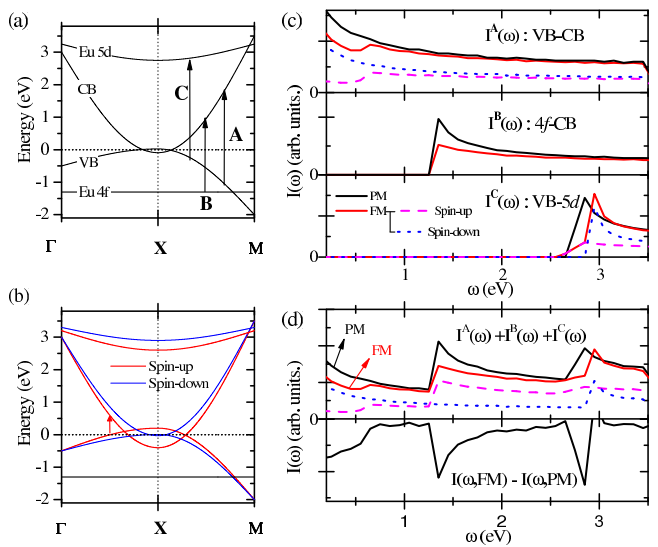}}
%\centerline{\includegraphics[width=4.5in,angle=-90]{fig3.eps}}%
\vspace*{0.2cm}%
%\centering\epsfig{file=fig3.eps,width=4cm,angle=-90}
%
\caption{(Color online) Schematic band structure of EuB$_{6}$ for
(a) PM and (b) FM states. In (a), the arrows A,B, and C indicate the
interband optical transitions. In (b), the small arrow indicates
onset of the A transition ($\uparrow$ channel). (c) Optical
absorption spectra I($\omega$) of the A,B, and C transitions
calculated for PM and FM. The FM curves are resolved into the
spin-up and spin-down components. (d) Total interband absorption
spectra
I($\omega$)=I$^{A}$($\omega$)+I$^{B}$($\omega$)+I$^{C}$($\omega$)
(upper panel) and the I($\omega$) difference between PM and FM
states (lower panel).} \label{fig:fig4}
\end{figure}

To gain insights into the $\sigma_{1}(\omega)$ weight transfer, we
start from the band calculation results by Massidda $et~al.$
\cite{massidda} and Kune{\v{s}} $et~al.$ \cite{kunes}. In
Fig.~\ref{fig:fig4}(a), the flat band at -1~eV represents the
localized Eu~4f$^{7}$ levels. The valence band (VB) and the
conduction band (CB) originate from B~2p and Eu~5d$_{x^2-y^2}$
orbital hybridization. The two bands cross the Fermi level at X
producing the semimetal hole and electron carriers. An empty Eu~5d
band is located at 3~eV.

Now let us consider possible optical interband transitions. The
arrows A, B, and C indicate dipole-allowed transitions
VB$\rightarrow$CB, Eu~4f$\rightarrow$CB, and VB$\rightarrow$Eu~5d
respectively. We will calculate their absorption spectra
I($\omega$)=2$\cdot\int M_{fi}(k)\delta(\omega-\Delta E(k)_{fi})dk$
where $\Delta E(k)_{fi}=E(k)_{f}-$$E(k)_{i}$ is the energy
difference between the final and initial state, and $M_{fi}(k)$ is
the transition matrix element. The prefactor 2 represents the
spin-up and down channels. In B, the \emph{i}-band electrons have
the same spin ($\uparrow$) and therefore only the $\uparrow$ channel
is activated. We assume that $M_{fi}(k)$ is constant for simplicity.
The calculated I($\omega$) for A,B, and C are shown
Fig.~\ref{fig:fig4}(c).

The band electrons interact with localized 4f moment through s-f
Kondo exchange coupling \cite{kunes}. In the FM phase, the bands
split into two spin-polarized sub-bands as a result. In CB the
spin-up (parallel to the Eu~4f spin) band is lowered in energy while
the spin-down (antiparallel) band shifts upwards. Eu~5d band splits
similarly. In contrast for the VB, the exchange integral has
opposite sign and the spin-bands shift to opposite directions.
According to Ref.~17, the split at X is $\sim$0.3~eV (CB) and 0.2~eV
(VB) and it decreases as $k\rightarrow \Gamma$ or $M$. In
Fig.~\ref{fig:fig4}(b), we reproduce the original LDA+U bands
approximately using simplified form $E(k)=E_0+W\cdot sin(ka/2)$
\cite{bands}.

% Figure 4
%

The band splitting brings about interesting changes in the Drude
conductivity and I($\omega$). For VB($\uparrow$), the upward band
shift leads to two effects; (i) the hole pocket expands (ii) the
number of the occupied state (E($k$)$<$E$_F$) decreases. (ii) occurs
at the expense of (i). (i) will increase the Drude conductivity. Due
to (ii) we can foretell that IB transition originating from this
band will decrease. For VB($\downarrow$), the band shifts oppositely
and so are the effects (i) and (ii). In CB($\uparrow$), the effects
are similar with VB($\uparrow$), i.e, (i) the electron pocket
expands. (ii) the empty states (E($k$)$>$E$_F$) decrease. On this
basis we calculate I($\omega$)= I($\omega,\uparrow$) +
I($\omega,\downarrow$) for FM. First consider A($\uparrow$)
transition where ($\emph{i}$, \emph{f}) = (VB($\uparrow$),
CB($\uparrow$)). In both bands $k$-space available for the
transition decreases due to (ii). I$^A$($\omega$,$\uparrow$) in
Fig.4 (c) shows that the transition is suppressed significantly at
$\omega<$ 0.5~eV. Along $X\rightarrow\Gamma$ note that the
transition has an onset at $k$ where E($k$)$_{i}=$E$_F$ as indicated
by the small arrow in Fig.~\ref{fig:fig4}(b). It has finite energy
$\sim$ 0.5~eV below which the transition is absent. In PM the onset
energy was almost zero. The I$^A$($\omega$,$\uparrow$) suppression
is due to this transition gap created by the band splittings. For
A($\downarrow$) channel the effects (ii) are opposite and
accordingly I($\omega$) will increase. However, the change is less
dramatic than in A($\uparrow$): note that I$^A$($\omega,\downarrow$)
is almost the same as that in PM, $\frac{1}{2}I^A$($\omega$,PM). In
total, I$^A$($\omega$)= I$^A$($\omega,\uparrow$) +
I$^A$($\omega,\downarrow$) is reduced  in FM. Similar I($\omega$)
suppression occur in B and C. In B, the \emph{i}-band Eu4f
($\uparrow$) remains without shift while the \emph{f}-band
CB($\uparrow$) shifts downward. Again by (ii), I$^B$($\omega$)
decreases. In C, I$^C$($\omega,\uparrow$) is suppressed due to the
\emph{i}-band shift similarly to that of A.

In Fig.~\ref{fig:fig4}(d), we show the total intensity spectra
I($\omega$)=I$^{A}$($\omega$)+I$^{B}$($\omega$)+I$^{C}$($\omega$)
for PM and FM and their difference
$\Delta$I($\omega$)=I($\omega$,FM)$-$I($\omega$,PM). The latter
shows excellent agreement with the $\Delta\sigma_{1}(\omega,T)$ data
in Fig.~\ref{fig:fig3}(a). The three-hump structure is reproduced.
The peak positions and the shapes agree remarkably well \cite{com2}.
To ensure the result we consider intermediate temperatures
5~K$<T<$15~K where the FM ordering is partial. Here the s-f band
splitting will be reduced. We repeat the calculation with smaller
splittings and find that I($\omega$) is consistent with the data,
i.e, the hump strength becomes weaker while the positions are
maintained.

In the Drude part, the hole and electron density change due to the
effect (i). It will alter the plasma frequency
$\omega_p^2=4{\pi}n/m_b$. We estimate the density n and the band
mass m$_b$ from the Fermi surface volume and the band dispersion,
respectively and calculate the total
$\omega_p^2$=[$\omega_p^2$($\uparrow$)
+$\omega_p^2$($\downarrow$)]$_{\emph{h}}$+[$\omega_p^2$($\uparrow$)+$\omega_p^2$($\downarrow$)]$_{\emph{e}}$,
the sum of the four contributions. We obtain
$\omega_{p,PM}\sim$~3500cm$^{-1}$ and
$\omega_{p,FM}^2$/$\omega_{p,PM}^2\sim$4.9. The $\omega_p^2$
enhancement comes mostly from the \emph{n} increase of the
\emph{h}($\uparrow$) and \emph{e}($\uparrow$) which overwhelms the
\emph{n} decrease of the ($\downarrow$) ones. The change of m$_b$ is
negligibly small. To compare with experiment, we calculate
$\omega_p^2$ from the sum rule $\omega_p^2
=\frac{120}{\pi}\int_{0^+}^{\omega_c}\sigma_1(\omega)d\omega$ with
$\omega_c$=0.3~eV. By taking $\sigma_1(\omega)$ at T=5~K (=FM) and
20~K (=PM), we find $\omega_{p,PM}\sim$~3300cm$^{-1}$ and
$\omega_{p,FM}^2$/$\omega_{p,PM}^2\sim$4.4. Again, the results are
in good agreement. It shows that the band splitting is the primary
source of the $\omega_{P}^{2}$ increase and consequently the
$\rho(T)$ drop at T$<$T$_{c}$.

The success of the IB and Drude analysis shows clearly that the band
diagram we employed represents the intrinsic EuB$_{6}$ electronic
structure. The multiple structure IB $\sigma_1(\omega)$ change stems
from the VB, CB, and Eu~5d band spin-splittings. It is particular
that VB and CB cross E$_{F}$ because their splittings remove
$\sigma_1(\omega)$ weight from the IB part and restore it into the
Drude part. We showed that in this way the Drude weight can come
from as high as 2.75~eV ($\gamma$), seeming unlikely remote energy.
If both VB and CB did not cross E$_{F}$ (insulator picture) or one
of them didn't, the observed $\sigma_1(\omega)$ change is not
produced. It is needed that both bands cross E$_{F}$ to explain the
multiple-structure optical weight transfer. It therefore supports
the semimetal band picture. Although our I($\omega)$ calculation is
rather primitive (k-dependence of $M_{ij}(k)$ is ignored and only
part of the whole BZ was taken into account), the analysis is
sufficient to support the conclusions. In the m*-change picture
\cite{hirsch99,hirsch} or the localization-delocaliztion picture
\cite{pereira} mentioned in the introduction, CB or VB change in
width or in the mobility-edge position. One needs to calculate
$\sigma_1(\omega)$ change for each case and compare with our result.
While rigorous analysis is not available at this point, it appears
that they both fail to explain the observed data.

% Figure 5
%
\begin{figure}[t]
%
%
% manuscript
%
\vspace*{0.0cm}\centerline{\includegraphics[width=1.9in,angle=0]{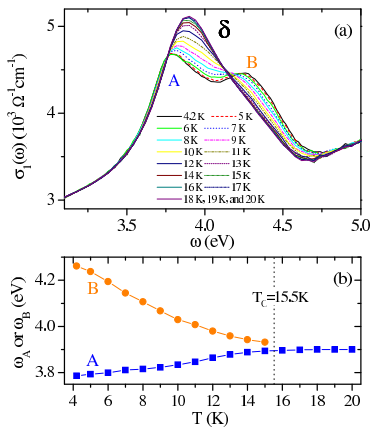}}
%\centerline{\includegraphics[width=4.5in]{fig2.eps}}%
\vspace*{0.0cm} \caption{(Color online) (a) Interband transition
$\delta$ and its T-dependent changes for T$<$20~K. (b) Peak
positions below the splitting at T$_{C}$=15.5~K. } \label{fig:fig5}
\end{figure}

In Fig.~\ref{fig:fig5}, we show the high energy peak $\delta$ at
4~eV. At T$<$T$_{c}$ it splits and the splitting reaches 0.48~eV at
T=4.2~K. This behavior is perhaps another result of the band
splitting. We associate $\delta$ with VB and Eu5d(II) band
\cite{fig5}. Here Eu5d(II) represents an empty band at 3~eV$-$5~eV
region (not shown in Fig.~\ref{fig:fig4}). Both Eu5d(II) and VB
exhibit van Hove singularities at $\Gamma$ and therefore VB
$\rightarrow$Eu5d(II) transition will contribute strong absorption
intensity like $\delta$. At $\Gamma$, Eu5d(II) is at +3.5~eV and it
splits by $\sim$0.5~eV in FM whereas VB is at -0.5~eV and remains
unsplitted. \cite{kunes}. Then the $\Gamma$-transition from VB to
Eu5d(II) will have 4~eV energy and 0.5~eV splitting. The $\delta$
splitting will be induced by magnetic field as well which may be
useful for UV magneto-optical application.

In summary we found that optical conductivity of EuB$_6$ exhibits
rich and novel spectral changes along with the FM transition: the
multiple-energy IB suppression (0.5, 1.4, and 2.75 eV), the
absorption peak splitting at 4~eV, and the Drude weight increase. We
showed that they are results of the semimetal band structure and the
Kondo coupling. This work provides new insights on the two long
standing unsolved issues of EuB$_6$.

We thank C.C. Homes for helping reflectivity measurement. This work
was supported by the KRF Grant No. -070-C00032 and the KOSEF through
CSCMR. The work at Toronto was supported by NSERC Discovery Grant
$\&$ Research Tools and Instrumentation grant.

%
%%%%%%%%%%%%%%%%%%%%%%%%%%%%%%%%%%%%%%%%%%%%%%%%%%%%%%%%%%%%%%%%%%%%%%%%%%%%%%
%
% References
%

\bibliography{teub6}

\begin{thebibliography}{20}
\expandafter\ifx\csname natexlab\endcsname\relax\def\natexlab#1{#1}\fi
\expandafter\ifx\csname bibnamefont\endcsname\relax
  \def\bibnamefont#1{#1}\fi
\expandafter\ifx\csname bibfnamefont\endcsname\relax
  \def\bibfnamefont#1{#1}\fi
\expandafter\ifx\csname citenamefont\endcsname\relax
  \def\citenamefont#1{#1}\fi
\expandafter\ifx\csname url\endcsname\relax
  \def\url#1{\texttt{#1}}\fi
\expandafter\ifx\csname urlprefix\endcsname\relax\def\urlprefix{URL }\fi
\providecommand{\bibinfo}[2]{#2}
\providecommand{\eprint}[2][]{\url{#2}}

\bibitem[{\citenamefont{Guy et~al.}(1980)\citenamefont{Guy, von Molnar,
  Etourneau, and Fisk}}]{Guy80}
\bibinfo{author}{\bibfnamefont{C.~N.} \bibnamefont{Guy}},
  \bibinfo{author}{\bibfnamefont{S.}~\bibnamefont{von Molnar}},
  \bibinfo{author}{\bibfnamefont{J.}~\bibnamefont{Etourneau}},
  \bibnamefont{and} \bibinfo{author}{\bibfnamefont{Z.}~\bibnamefont{Fisk}},
  \bibinfo{journal}{Solid State Commun.} \textbf{\bibinfo{volume}{33}},
  \bibinfo{pages}{1055} (\bibinfo{year}{1980}).

\bibitem[{\citenamefont{Degiorgi et~al.}(1997)\citenamefont{Degiorgi, Felder,
  Ott, Sarrao, and Fisk}}]{degiorgi}
\bibinfo{author}{\bibfnamefont{L.}~\bibnamefont{Degiorgi}},
  \bibinfo{author}{\bibfnamefont{E.}~\bibnamefont{Felder}},
  \bibinfo{author}{\bibfnamefont{H.~R.} \bibnamefont{Ott}},
  \bibinfo{author}{\bibfnamefont{J.~L.} \bibnamefont{Sarrao}},
  \bibnamefont{and} \bibinfo{author}{\bibfnamefont{Z.}~\bibnamefont{Fisk}},
  \bibinfo{journal}{Phys. Rev. Lett.} \textbf{\bibinfo{volume}{79}},
  \bibinfo{pages}{5134} (\bibinfo{year}{1997}).

\bibitem[{cmr()}]{cmr}
\bibinfo{note}{Colossal Magnetoresistance Oxides, edited by Y. Tokura (Gordon
  $\&$ Breach, New York, 2000).}

\bibitem[{\citenamefont{Kreissl and Nolting}(2005)}]{kreissl}
\bibinfo{author}{\bibfnamefont{M.}~\bibnamefont{Kreissl}} \bibnamefont{and}
  \bibinfo{author}{\bibfnamefont{W.}~\bibnamefont{Nolting}},
  \bibinfo{journal}{Phys. Rev. B} \textbf{\bibinfo{volume}{72}},
  \bibinfo{pages}{245117} (\bibinfo{year}{2005}).

\bibitem[{\citenamefont{Hirsch}(1999)}]{hirsch99}
\bibinfo{author}{\bibfnamefont{J.~E.} \bibnamefont{Hirsch}},
  \bibinfo{journal}{Phys. Rev. B} \textbf{\bibinfo{volume}{59}},
  \bibinfo{pages}{436} (\bibinfo{year}{1999}).

\bibitem[{\citenamefont{Hirsch}(2000)}]{hirsch}
\bibinfo{author}{\bibfnamefont{J.~E.} \bibnamefont{Hirsch}},
  \bibinfo{journal}{Phys. Rev. B} \textbf{\bibinfo{volume}{62}},
  \bibinfo{pages}{14131} (\bibinfo{year}{2000}).

\bibitem[{\citenamefont{Pereira et~al.}(2004)\citenamefont{Pereira, dos Santos,
  Castro, and Neto}}]{pereira}
\bibinfo{author}{\bibfnamefont{V.~M.} \bibnamefont{Pereira}},
  \bibinfo{author}{\bibfnamefont{J.~B.~L.} \bibnamefont{dos Santos}},
  \bibinfo{author}{\bibfnamefont{E.~V.} \bibnamefont{Castro}},
  \bibnamefont{and} \bibinfo{author}{\bibfnamefont{A.~H.~C.}
  \bibnamefont{Neto}}, \bibinfo{journal}{Phys. Rev. Lett}
  \textbf{\bibinfo{volume}{93}}, \bibinfo{pages}{147202}
  (\bibinfo{year}{2004}).

\bibitem[{\citenamefont{Goodrich et~al.}(1998)\citenamefont{Goodrich, Harrison,
  Vuillemin, Teklu, Hall, Fisk, Young, and Sarrao}}]{goodrich}
\bibinfo{author}{\bibfnamefont{R.~G.} \bibnamefont{Goodrich}},
  \bibinfo{author}{\bibfnamefont{N.}~\bibnamefont{Harrison}},
  \bibinfo{author}{\bibfnamefont{J.~J.} \bibnamefont{Vuillemin}},
  \bibinfo{author}{\bibfnamefont{A.}~\bibnamefont{Teklu}},
  \bibinfo{author}{\bibfnamefont{D.~W.} \bibnamefont{Hall}},
  \bibinfo{author}{\bibfnamefont{Z.}~\bibnamefont{Fisk}},
  \bibinfo{author}{\bibfnamefont{D.}~\bibnamefont{Young}}, \bibnamefont{and}
  \bibinfo{author}{\bibfnamefont{J.}~\bibnamefont{Sarrao}},
  \bibinfo{journal}{Phys. Rev. B} \textbf{\bibinfo{volume}{58}},
  \bibinfo{pages}{14896} (\bibinfo{year}{1998}).

\bibitem[{\citenamefont{Aronson et~al.}(1999)\citenamefont{Aronson, Sarrao,
  Fisk, Whitton, and Brandt}}]{aronson}
\bibinfo{author}{\bibfnamefont{M.~C.} \bibnamefont{Aronson}},
  \bibinfo{author}{\bibfnamefont{J.~L.} \bibnamefont{Sarrao}},
  \bibinfo{author}{\bibfnamefont{Z.}~\bibnamefont{Fisk}},
  \bibinfo{author}{\bibfnamefont{M.}~\bibnamefont{Whitton}}, \bibnamefont{and}
  \bibinfo{author}{\bibfnamefont{B.~L.} \bibnamefont{Brandt}},
  \bibinfo{journal}{Phys. Rev. B} \textbf{\bibinfo{volume}{59}},
  \bibinfo{pages}{4720} (\bibinfo{year}{1999}).

\bibitem[{\citenamefont{Denlinger et~al.}(2002)\citenamefont{Denlinger, Clack,
  Allen, Gweon, Poirier, Olson, Sarrao, Bianchi, and Fisk}}]{denlinger}
\bibinfo{author}{\bibfnamefont{J.~D.} \bibnamefont{Denlinger}},
  \bibinfo{author}{\bibfnamefont{J.~A.} \bibnamefont{Clack}},
  \bibinfo{author}{\bibfnamefont{J.~W.} \bibnamefont{Allen}},
  \bibinfo{author}{\bibfnamefont{G.-H.} \bibnamefont{Gweon}},
  \bibinfo{author}{\bibfnamefont{D.~M.} \bibnamefont{Poirier}},
  \bibinfo{author}{\bibfnamefont{C.~G.} \bibnamefont{Olson}},
  \bibinfo{author}{\bibfnamefont{J.~L.} \bibnamefont{Sarrao}},
  \bibinfo{author}{\bibfnamefont{A.~D.} \bibnamefont{Bianchi}},
  \bibnamefont{and} \bibinfo{author}{\bibfnamefont{Z.}~\bibnamefont{Fisk}},
  \bibinfo{journal}{Phys. Rev. Lett.} \textbf{\bibinfo{volume}{89}}
  (\bibinfo{year}{2002}).

\bibitem[{\citenamefont{Rhyee et~al.}(2003{\natexlab{a}})\citenamefont{Rhyee,
  Cho, and Ri}}]{rhyee2}
\bibinfo{author}{\bibfnamefont{J.-S.} \bibnamefont{Rhyee}},
  \bibinfo{author}{\bibfnamefont{B.~K.} \bibnamefont{Cho}}, \bibnamefont{and}
  \bibinfo{author}{\bibfnamefont{H.-.~C.} \bibnamefont{Ri}},
  \bibinfo{journal}{Phys. Rev. B} \textbf{\bibinfo{volume}{67}},
  \bibinfo{pages}{125102} (\bibinfo{year}{2003}{\natexlab{a}}).

\bibitem[{\citenamefont{Rhyee et~al.}(2003{\natexlab{b}})\citenamefont{Rhyee,
  Oh, Cho, Kim, and Jung}}]{rhyee3}
\bibinfo{author}{\bibfnamefont{J.-S.} \bibnamefont{Rhyee}},
  \bibinfo{author}{\bibfnamefont{B.~H.} \bibnamefont{Oh}},
  \bibinfo{author}{\bibfnamefont{B.~K.} \bibnamefont{Cho}},
  \bibinfo{author}{\bibfnamefont{H.~C.} \bibnamefont{Kim}}, \bibnamefont{and}
  \bibinfo{author}{\bibfnamefont{M.~H.} \bibnamefont{Jung}},
  \bibinfo{journal}{Phys. Rev. B} \textbf{\bibinfo{volume}{67}},
  \bibinfo{pages}{212407} (\bibinfo{year}{2003}{\natexlab{b}}).

\bibitem[{\citenamefont{Homes et~al.}(1993)\citenamefont{Homes, Reedyk,
  Crandles, and Timusk}}]{homes93}
\bibinfo{author}{\bibfnamefont{C.~C.} \bibnamefont{Homes}},
  \bibinfo{author}{\bibfnamefont{M.}~\bibnamefont{Reedyk}},
  \bibinfo{author}{\bibfnamefont{D.~A.} \bibnamefont{Crandles}},
  \bibnamefont{and} \bibinfo{author}{\bibfnamefont{T.}~\bibnamefont{Timusk}},
  \bibinfo{journal}{Appl. Opt.} \textbf{\bibinfo{volume}{32}},
  \bibinfo{pages}{2976} (\bibinfo{year}{1993}).

\bibitem[{\citenamefont{Kuzmenko}(2005)}]{kuzmenko05}
\bibinfo{author}{\bibfnamefont{A.~B.} \bibnamefont{Kuzmenko}},
  \bibinfo{journal}{Rev. Sci. Instrum.} \textbf{\bibinfo{volume}{76}}
  (\bibinfo{year}{2005}).

\bibitem[{com({\natexlab{a}})}]{com1}
\bibinfo{note}{The change of $\alpha$ arises from $R(\omega)$ above the plasma
  edge (at 0.1-0.6~eV) where $R(\omega)$ decreases as T decreases.}

\bibitem[{\citenamefont{Massidda et~al.}(1997)\citenamefont{Massidda,
  Continenza, Pascale, and Monnier}}]{massidda}
\bibinfo{author}{\bibfnamefont{S.}~\bibnamefont{Massidda}},
  \bibinfo{author}{\bibfnamefont{A.}~\bibnamefont{Continenza}},
  \bibinfo{author}{\bibfnamefont{T.~M.~D.} \bibnamefont{Pascale}},
  \bibnamefont{and} \bibinfo{author}{\bibfnamefont{R.}~\bibnamefont{Monnier}},
  \bibinfo{journal}{Z. Phys. B} \textbf{\bibinfo{volume}{102}},
  \bibinfo{pages}{83} (\bibinfo{year}{1997}).

\bibitem[{\citenamefont{Kune{\v{s}} and Pickett}(2004)}]{kunes}
\bibinfo{author}{\bibfnamefont{J.}~\bibnamefont{Kune{\v{s}}}} \bibnamefont{and}
  \bibinfo{author}{\bibfnamefont{W.~E.} \bibnamefont{Pickett}},
  \bibinfo{journal}{Phys. Rev. B} \textbf{\bibinfo{volume}{69}},
  \bibinfo{pages}{165111} (\bibinfo{year}{2004}).

\bibitem[{ban()}]{bands}
\bibinfo{note}{($E_0,W_{\uparrow},W_{\downarrow}$)=(3,-3.4,-3.03)$_{\rm
  X-\Gamma}$ and (3.5,-3.9,-3.53)$_{\rm X-M}$ for CB, (0.5,0.7,0.505)$_{\rm
  X-\Gamma}$ and (-2,2.2,2.005)$_{\rm X-M}$ for VB, and (3.25,-0.6,-0.4) for
  Eu~5d. To plot the PM bands, we took spin-average of the FM bands.}

\bibitem[{com({\natexlab{b}})}]{com2}
\bibinfo{note}{The Drude tail overlaps with the $\alpha$-hump. The suppression
  of -$\Delta\sigma_1(\omega,T)$ at $\omega<$0.5~eV is due to the T-dependent
  Drude part which is not accounted for in the I($\omega$) calculation.}

\bibitem[{fig()}]{fig5}
\bibinfo{note}{From Eu~5d orbital, total 5 empty bands are derived where the
  +3eV one is the lowest-lying branch. We refer the next branch as Eu5d(II).}

\end{thebibliography}

\end{document}